\documentclass[letterpaper,10 pt,conference]{ieeeconf}
\IEEEoverridecommandlockouts
\usepackage[utf8]{inputenc}
\usepackage{amsmath}
\usepackage{mathrsfs}
\usepackage{amssymb}
\usepackage{color}
\usepackage{dsfont}
\usepackage{cite}
\usepackage{comment}
\usepackage{float}
\usepackage{graphicx}
\usepackage[font = footnotesize]{caption}
\usepackage[font = footnotesize]{subcaption}
\usepackage{mathStyle}
\usepackage{color}
\usepackage{xcolor}
\usepackage{float}
\usepackage{tikz}
\usepackage{pgfplots}
\usepackage{booktabs}
\usepackage{tabularx}
\usepackage{multirow}
\usepackage{bm}
\usepackage{makecell}
\usepackage{indentfirst}
\newcolumntype{C}[1]{>{\centering\let\newline\\\arraybackslash\hspace{0pt}}m{#1}}

\usepackage{color, colortbl}
\definecolor{Yellow}{rgb}{1,1,0}

\newcommand{\argmin}{\mathop{\mathrm{argmin}}\limits}

\usepackage[linesnumbered,ruled,vlined]{algorithm2e}
\SetKwInOut{KwParam}{Parameters}
\let\oldnl\nl
\newcommand{\nonl}{\renewcommand{\nl}{\let\nl\oldnl}}

\usepackage{hhline}

\usetikzlibrary{calc}
\usetikzlibrary{shapes, arrows}
\usetikzlibrary{positioning}
\usepackage{circuitikz}
\usetikzlibrary{calc,patterns,decorations.pathmorphing,decorations.markings,positioning}

\usetikzlibrary{shapes,arrows,calc,positioning}
\tikzstyle{bigblock} = [draw, fill=blue!20, rectangle, 
    minimum height=6em, minimum width=8em]
\tikzstyle{medblock} = [draw, fill=blue!20, rectangle, 
    minimum height=4em, minimum width=4em]    
\tikzstyle{mux} = [draw, fill=black!20, rectangle, 
    minimum height=5em, minimum width=0.1em]    
\tikzstyle{smallblock} = [draw, fill=blue!20, rectangle, 
    minimum height=2.5em, minimum width=4em]
\tikzstyle{sum} = [draw, fill=blue!20, circle, node distance=1cm]
\tikzstyle{signal} = [coordinate]
\tikzstyle{pinstyle} = [pin edge={to-,thin,black}]
\tikzstyle{block} = [draw, fill=blue!20, rectangle, 
    minimum height=3em, minimum width=6em]
\tikzstyle{blockS} = [draw, fill=blue!20, rectangle, 
    minimum height=3em, minimum width=4em]  
\tikzstyle{sum} = [draw, fill=blue!20, circle, node distance=1.5cm]
\tikzstyle{gain} = [draw, fill=blue!20, regular polygon, regular polygon sides = 3, node distance=1.25cm, shape border rotate = -90]
\tikzstyle{mult} = [draw, fill=blue!20, circle, inner sep=0pt, minimum size=0.2cm,]
\tikzstyle{input} = [coordinate]
\tikzstyle{output} = [coordinate]
\tikzstyle{spring}=[thick,decorate,decoration={zigzag,pre length=0.3cm,post length=0.3cm,segment length=6}]
\tikzstyle{damper}=[thick,decoration={markings,  
  mark connection node=dmp,
  mark=at position 0.5 with 
  {
    \node (dmp) [thick,inner sep=0pt,transform shape,rotate=-90,minimum width=15pt,minimum height=3pt,draw=none] {};
    \draw [thick] ($(dmp.north east)+(2pt,0)$) -- (dmp.south east) -- (dmp.south west) -- ($(dmp.north west)+(2pt,0)$);
    \draw [thick] ($(dmp.north)+(0,-5pt)$) -- ($(dmp.north)+(0,5pt)$);
  }
}, decorate]
\tikzstyle{ground}=[fill,pattern=north east lines,draw=none,minimum width=0.75cm,minimum height=0.3cm]

\newcounter{example}

\usepackage{hyperref}
\usepackage{xcolor}
\hypersetup{
    colorlinks,
    linkcolor={blue!100!black},
    citecolor={blue!50!black},
    urlcolor={blue!80!black}
}

\title{Predictive Control Barrier Functions for \\ Discrete-Time Linear Systems with Unmodeled Delays}

\author{
Juan Augusto Paredes Salazar,
James Usevitch, 
and
Ankit Goel
\thanks{James Usevitch is with the Department of Electrical Engineering, BYU. {\tt\small james\_usevitch@byu.edu}}%
\thanks{Juan Augusto Paredes Salazar and Ankit Goel are with the Department of Mechanical Engineering, University of Maryland, Baltimore County,1000 Hilltop Circle, Baltimore, MD 21250. {\tt\small japarede, ankgoel@umbc.edu }}%
}

\begin{document}

\maketitle

\begin{abstract}
This paper introduces a predictive control barrier function (PCBF) framework for enforcing state constraints in discrete-time systems with unknown relative degree, which can be caused by input delays or unmodeled input dynamics. 
Existing discrete-time CBF formulations typically require the construction of auxiliary barrier functions when the relative degree is greater than one, which complicates implementation and may yield conservative safe sets. 
The proposed PCBF framework addresses this challenge by extending the prediction horizon to construct a CBF for an associated system with relative degree one. 
As a result, the superlevel set of the PCBF coincides with the safe set, simplifying constraint enforcement and eliminating the need for auxiliary functions. 
The effectiveness of the proposed method is demonstrated on a discrete-time double integrator with input delay and a bicopter system with position constraints.
\end{abstract}


\section{Introduction} \label{sec:intro}

Control barrier functions (CBFs) have emerged as a powerful framework for enforcing safety in control systems by guaranteeing the forward invariance of a prescribed safe set.
CBFs are grounded in Nagumo’s forward invariance theorem, which characterizes the conditions under which trajectories of an ordinary differential equation remain inside a given set \cite{menner2024translation}.
By ensuring that Nagumo’s condition holds, CBFs enforce state constraints and yield a safety filter that modifies the nominal control input to maintain constraint satisfaction.
Several variants of CBFs have been proposed in the literature, including zeroing CBFs, higher-order CBFs, and related formulations 
%
%
\cite{ames2019,tan2021,garg2024}. 
These methods provide a systematic framework for constraint satisfaction, often through the solution of constrained optimization problems.


However, most CBF formulations are developed in continuous time, which necessitates discretization for practical implementation in digital control systems. 
Since safety constraints are only evaluated at discrete sampling instants, constraint violations may occur between updates. 
Moreover, discrete enforcement of CBF conditions can lead to aggressive corrective actions when a constraint is about to be violated. These limitations have motivated several extensions, including sampled-data CBFs 
\cite{breeden2021,usevitch2022,tan2025},
robust formulations that account for inter-sample effects 
%
%
\cite{jankovic2018,zhang2022,bahati2024,alan2025},
and event-triggered or self-triggered update mechanisms 
%
%
\cite{yang2019,taylor2020,sabouni2024}.
To directly address discrete dynamics, several recent works have focused on discrete-time CBF (DT-CBF) formulations \cite{agrawal2017,xiong2022,zheng2024}.
Given that model predictive control (MPC) is inherently implemented in discrete time, DT-CBFs have been widely employed to enhance constraint enforcement in MPC frameworks \cite{ma2021,zeng2021_2,wabersich2022,liu2023_MPC_CBF,hall2023,liu2025,priess2025}.

%
%

%
%

A central challenge in designing safety filters and ensuring the forward invariance of a desired safe set lies in the relative degree of the constraint function with respect to the control input.
When the relative degree is greater than one, it is necessary to construct a sequence of auxiliary barrier functions equal in number to the relative degree, with forward invariance guaranteed only for the intersection of the superlevel sets of these auxiliary functions.
However, the resulting safe set is often challenging to compute explicitly and may be overly conservative or impractical for control design.

This paper is focused on the problem of designing safety filters for discrete-time systems with unknown relative degree arising from input delays \cite{jankovic2018_delay,kiss2023} or unmodeled input dynamics \cite{seiler2021,quan2023,alan2025}.
The main contribution is the development of a predictive CBF (PCBF) framework.
By extending the prediction horizon, the proposed approach constructs a CBF for an associated system with relative degree one, thereby eliminating the need to design auxiliary barrier functions that are typically required in the high relative-degree case.
Furthermore, reducing the relative degree to one ensures that the superlevel set of the control barrier function directly defines the safe set.

The contents of this paper are as follows.
Section \ref{sec:problem} briefly reviews the sampled-data feedback control problem for a continuous-time dynamic system and introduces the CBF operation in the control loop.
Section \ref{sec:cbf} presents the formulation of a CBF for DT systems with high relative degree.
Section \ref{sec:pcbf} presents the formulation of the PCBF framework, which extends the formulation in Section \ref{sec:cbf} to the case where the DT system dynamics are linear, allowing for the evaluation of CBF conditions over a prediction horizon.
Section \ref{sec:simulations} presents examples that illustrate the performance of the proposed PCBF algorithm and its effectiveness at enforcing state constraints.
Finally, the paper concludes with a summary in Section \ref{sec:conclusions}.


\section{Problem Formulation}\label{sec:problem}

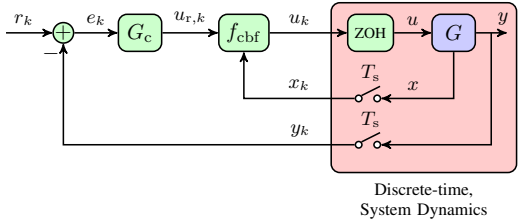
\begin{figure} [h!]
    \centering
    \resizebox{0.8\columnwidth}{!}
    {%
    \begin{tikzpicture}[>={stealth'}, line width = 0.25mm]

    \node [input, name=ref]{};
    \node [smallblock, fill=red!20, rounded corners, below right = -0.5 cm and 5.25cm of ref , minimum height = 2.75cm , minimum width = 3cm] (system_dt) {};
    \node[below] at (system_dt.south) {\footnotesize $\begin{array}{c} \mbox{Discrete-time,} \\  \mbox{System Dynamics} \end{array}$};
    \node [sum, fill=green!20, right =0.75cm of ref] (sum2) {};
    \node[draw = white] at (sum2.center) {$+$};
    \node [smallblock, fill=green!20, rounded corners, right = 0.7cm of sum2 , minimum height = 0.6cm , minimum width = 0.7cm] (controller) {$G_\rmc$};
    \node [smallblock, fill=green!20, rounded corners, right = 0.9cm of controller , minimum height = 0.6cm , minimum width = 0.7cm] (cbf) {$f_{\rm cbf}$};
    \node [smallblock,  fill=green!20, rounded corners, right = 1.25cm of cbf, minimum height = 0.6cm , minimum width = 0.5cm] (DA) {\scriptsize ZOH};
    \node [smallblock, rounded corners, right = 0.6cm of DA, minimum height = 0.6cm , minimum width = 0.7cm] (system) {$G$};
    \node [output, right = 0.5cm of system] (output) {};
    \node [input, below right = 0.75cm and 0.4cm of system] (midpoint2) {};

    \node [input, below right = 1.5cm and 0.4cm of system] (midpoint) {};
    
    \draw [->] (controller) -- node [above] {\small $u_{\rmr, k}$} (cbf);
    \draw [->] (cbf) -- node [above, xshift = -0.4em] {\small $u_k$} (DA);
    \draw [->] (DA) -- node [above] (du) {\small$u$} (system);

    \node[circle,draw=black, fill=white, inner sep=0pt,minimum size=3pt] (rc12) at ([xshift=-2.0cm]midpoint2) {};
    \node[circle,draw=black, fill=white, inner sep=0pt,minimum size=3pt] (rc22) at ([xshift=-2.3cm]midpoint2) {};
    \draw [-] (rc22.north east) --node[below,yshift=.55cm]{\small $T_\rms$} ([xshift=.3cm,yshift=.15cm]rc22.north east) {};
    
    \node[circle,draw=black, fill=white, inner sep=0pt,minimum size=3pt] (rc11) at ([xshift=-2.0cm]midpoint) {};
    \node[circle,draw=black, fill=white, inner sep=0pt,minimum size=3pt] (rc21) at ([xshift=-2.3cm]midpoint) {};
    \draw [-] (rc21.north east) --node[below,yshift=.55cm]{\small $T_\rms$} ([xshift=.3cm,yshift=.15cm]rc21.north east) {};
    
    \draw [->] (system) -- node [name=y, near end]{} node [very near end, above] {\small $y$}(output);
    \draw [->] (system.south) |- node [near end, above, xshift = -0.2 em] {\small $x$}  (rc12.east);
    \draw [->](rc22.west) -|  node [near start, above, xshift = 0.0 em] {\small $x_k$} (cbf.south);
    
    \draw [->] (ref.east) -- node [near start, above,xshift = 0.1cm] {\small $r_k$}  node [near end, above] {} (sum2);
    \draw [->] (y.west) |- (rc11.east);
    \draw [->] (rc21) -| node [very near start, above, xshift = 0.3cm] {\small $y_k$} node [very near end, xshift = -0.2cm, yshift = 0.25cm] {$-$} (sum2.south);
    \draw [->] (sum2.east) -- node [above] {\small $e_k$} (controller.west);
    
    \end{tikzpicture}
    }  
    \caption{Sampled data implementation of discrete-time controller $G_\rmc$ applied to a continuous-time system $G$ with input $u,$ state $x,$ and output $y.$
    The function $f_{\rm cbf}$ implements the CBF by modifying the output of $G_\rmc$ for state constraint enforcement. 
    All sample-and-hold operations are synchronous.}
    \label{fig:CT_cbf_blk_diag}
\end{figure}

To reflect the practical implementation of digital controllers for physical systems, we consider continuous-time dynamics under sampled-data control using a discrete-time controller.
In particular, we consider the control architecture shown in Figure \ref{fig:CT_cbf_blk_diag}, where $G$ is the target continuous-time system, and $u,$ $x,$ and $y$ are the control input, internal state, and the output of $G,$ respectively.
The state $x$ and the output $y$ are sampled to produce the sampled state measurement $x_k,$ and the sampled output measurement $y_k,$ respectively, which, for all $k \ge 0,$ are given by $x_k \isdef x(k T_\rms)$ and $y_k \isdef y(k T_\rms),$ where $T_\rms > 0$ is the sample time.

The discrete-time controller is denoted by $G_\rmc.$ 
The input to $G_\rmc$ is the reference error $e_k \isdef r_k - y_k,$ where $r_k$ is the reference signal, and its output is the requested discrete-time control input $u_{\rmr, k}.$

The function $f_{\rm cbf}$ implements the safety filter synthesized by the CBF framework for state constraint enforcement. 
The inputs to $f_{\rm cbf}$ are the requested discrete-time control input $u_{\rmr, k}$ and the sampled state measurement $x_k,$ and its output is the discrete-time control input $u_k.$
Then, the continuous-time control signal $u$ applied to the system $G$ is generated by applying a zero-order-hold operation to $u_k,$ that is, for all $k \ge 0,$ and, for all $t \in [k T_\rms, (k+1) T_\rms),$ $u(t) = u_k.$

The controller $G_\rmc$ is designed so that $y_k$ follows $r_k,$ such that $\sum_{k = 0}^{\infty} \Vert r_k - y_k\Vert = \sum_{k = 0}^{\infty} \Vert e_k\Vert$ is minimized.
Next, let $\SC$ be a subset such that all state constraints are satisfied at step $k$ if and only if $x_k \in \SC.$
Under the assumption that $x_0 \in \SC,$ $f_{\rm cbf}$ is designed to modify $u_{\rmr, k}$ so that $x_k \in \SC$ for all $k > 0.$
%

\section{Review of Control Barrier Function for High Relative-Degree Discrete-Time Systems} \label{sec:cbf}
This section provides a brief review of the design of safety filters in discrete-time systems where the relative degree of the constraint functions with respect to the input is greater than one. 
The construction follows the procedure described in \cite{xiong2022}.
Consider the discrete-time dynamic system
\begin{align}
    x_{k+1} = f(x_k,u_k),
    \label{eq:gen_sys}
\end{align}
where $k \ge0$ is the time step, $x_k \in \BBR^n$ is the state, and $u_k \in \BBR^m$ is the control input.
Consider the control barrier function $h \colon \BBR^n \to \BBR^p$ and the corresponding desired safe set 
\begin{align}
    \SC_\rms
        \isdef 
            \{ x \in \BBR^n \colon h(x) \geq 0 \}.
\end{align}
Note that the function $h$ encodes $p$ constraints, and $h(x_k) \geq 0$ implies that $x_k$ satisfies all constraints. 


For $i =1,\ldots, p$, let $\rho_i$ denote the relative degree of $h_i(x)$ with respect to $u.$
Recall that the relative degree of $p(x_k)$ with respect to $u_k$ is the smallest integer $m$ such that $p(x_{k+m})$ depends explicitly on $u_k.$
For $i =1,\ldots, p$, define
\begin{align}
    \psi_{0,i}(x_k) 
        &\isdef
            h_i(x_k),
    \\ 
        & \ \vdots \nn
    \\
     \psi_{\rho_i-1,i}(x_k) 
        &\isdef
            \Delta [\psi_{\rho_i-2,i}(x_{k}) ]
            +
            \alpha (\psi_{\rho_i-2,i}(x_k)),
    \\
    \psi_{\rho_i,i}(x_k,u_k) 
        &\isdef
            \Delta [\psi_{\rho_i-1,i}(x_{k}) ]
            +
            \alpha (\psi_{\rho_i-1,i}(x_k)),
\end{align}
where
$\alpha:\BBR \to \BBR$ is a class $\SK$ function such that, for all $r > 0,$ $\alpha(r) < r$, and
$\Delta [p(x_{k}) ] \isdef p(x_{k+1})-p(x_k).$
Next, define the corresponding sets 
\begin{align}
    \SC_{0,i} 
        &\isdef 
            \{ x \in \BBR^n \colon \psi_{0,i}(x) \geq 0 \}, \\
        & \ \vdots \nn
    \\
    \SC_{\rho,i} 
        &\isdef 
            \{ x \in \BBR^n \colon \psi_{\rho,i}(x,u) \geq 0 \}.
\end{align}
Finally, define
\begin{align}
    \SC \isdef \SC_{0,1} \cap \SC_{1,1} \cdots \cap \SC_{\rho_p,p}.
\end{align}
\begin{theo}
    \label{theo:CBF_theorem}    
    If $x_0 \in \SC,$ then, for all $k \geq 0,$ $x_k \in \SC. $
\end{theo}
\textbf{Proof.}
See \cite{xiong2022}.
$\hfill \blacksquare$

\begin{remark}
    In general, the class $\SK$ functions $\alpha$ for each $\psi_{j,i}$ can be chosen differently.
\end{remark}

The safety filter that enforces the constraint $h_i(x_k)>0$ is obtained by setting $\psi_{\rho_i,i}(x_k,u_k) \geq 0.$
When multiple constraints are present in a system, and their relative degrees with respect to the input are not identical, the design of the safety filter and the corresponding forward-invariant set becomes a nontrivial task. 
As described above, a chain of auxiliary functions must be introduced to synthesize the safety filter. Furthermore, when the relative degrees of each constraint function differ, each constraint must be addressed separately. 

\subsection{Relative Degree = 1}
\label{sec:RD1}
Consider a control barrier function $h(x) \in \BBR^p$ such that each element of $h(x_k)$ has relative degree one with respect to the input $u_k$. 
Then, the safety filter is constructed as follows. 
Define
\begin{align}
    \psi_{0}(x_k) 
        &\isdef
            h(x_k),
    \\
    \psi_{1}(x_k, u_k) 
        &\isdef
            \Delta [\psi_{0}(x_{k}) ]
            +
            \lambda \psi_{0}(x_k)
        \nn \\
        &=
            \psi_{0}(x_{k+1}) 
            +
            (\lambda-1) \psi_{0}(x_k),
\end{align}
where $\lambda \in (0,1).$
Note that $\SC \isdef \{ x \in \BBR^n \colon h(x) \geq 0 \}.$
It follows from Theorem \ref{theo:CBF_theorem} that if $x_0 \in \SC$ and, for each $k,$ $u_k$ satisfies $\psi_{1}(x_k, u_k) \geq 0 ,$ then, $x_k \in \SC.$

Note that the condition $\psi_{1}(x_k, u_k) \geq 0$ is equivalent to
\begin{align}
    h(x_{k+1})  \geq (1-\lambda) h(x_k).
\end{align}

\begin{prop}
Let $k \ge 0$ and suppose that $ h(x_0) \ge 0$ and
\begin{align}
    h(x_{k+1}) \ge \gamma h(x_k), \label{eq:cbf_dt_v2}
\end{align}
where $\gamma \in (0, 1).$
Then, $h(x_k) \ge 0.$
\end{prop}

{\bf Proof:} Define $\nu_k \isdef h(x_k).$
It follows from \eqref{eq:cbf_dt_v2} that
$\nu_{k+1} \ge \gamma \nu_k,$
which implies by recursion that $h(x_k) = \nu_{k} \ge \gamma^k \nu_0 \ge 0.$
\hfill $\square$






\subsection{Convex Polytopic Constraints}

Convex polytopic constraints (CPC) arise when the admissible states, inputs, or parameters of a system are required to remain within a convex polytope.
In control applications, polytopic constraints can be used to model position and velocity bounds, actuator saturation limits, and regions of safe operation.
In particular, a polytopic constraint is a set of the form
\begin{align}
    \mathcal{P} = \{x \in \mathbb{R}^n : A x \leq b\},
\end{align}
which is the intersection of a finite number of half-spaces. 
For example, consider the position and velocity constraints 
\begin{align}
    p_{\min} \leq p \leq p_{\max}, \quad
    v_{\min} \leq v \leq v_{\max},
\end{align}
which can be written in polytopic form as
\begin{align}
    \matl
        1 & 0 \\ -1 & 0 \\ 0 & 1 \\ 0 & -1
    \matr
    \matl 
        p \\ v
    \matr
    \leq
    \matl
        x_{\max} \\ -x_{\min} \\ v_{\max} \\ -v_{\min}
    \matr.
\end{align}
Note that all polytopic constraints of the form $Ax\leq b$ are not necessarily convex.

\section{Predictive Control Barrier Function}
\label{sec:pcbf}
This section introduces the formulation of the predictive control barrier function. 
The central idea is that increasing the prediction horizon effectively reduces the relative degree of the constraints with respect to the input, thereby simplifying the synthesis and implementation of the safety filter.  
For example, consider a system with relative degree three. 
If the discrete-time dynamics are reformulated using a time step greater than three, it is easy to show that the relative degree of the resulting system is one.


As described in Section \ref{sec:RD1}, when the relative degree is one, control barrier function constraints reduce to a single inequality, which simplifies the safety filter design.
The safe set is simply the level set of $h(x)=0,$ which is forward invariant under any control satisfying the CBF condition. 
Moreover, multiple constraints are handled simultaneously in one single inequality.

For example, consider the discrete-time system \eqref{eq:gen_sys} and a scalar function $h(x)$ whose relative degree with respect to the input $u$ is $\rho.$
Define the horizon $\ell_\rmh>\rho$ and consider the system
\begin{align}
    \chi_{k+1} = f^{\ell_\rmh} (\chi_k,u_k), 
    \label{eq:chi_dyn}
\end{align}
where $$f^{\ell_\rmh} \isdef f(f( \ldots f(x,u),u),\ldots), u),$$ with the map $f$ applied $\ell_\rmh$ times.
Then, the relative degree of $h(\chi)$ with respect to $u$ is reduced to one. 
Note that the effective time step of \eqref{eq:chi_dyn} is $\ell_\rmh$ times the time step of \eqref{eq:gen_sys}.
However, we emphasize that, since the control $u$ is assumed to be constant over the $\ell_\rmh$ steps, the dynamics given by \eqref{eq:chi_dyn} is not equivalent to the dynamics given by \eqref{eq:gen_sys}.

Several control techniques have been developed based on the assumption of constant inputs over the prediction horizon, such as reference–governor schemes \cite{kolmanovsky2014,garone2017,garone2018},
%
move–blocking strategies in receding–horizon control, which parameterize $u$ as piecewise–constant over a few blocks 
%
%
\cite{cagienard2007,shekhar2015,makarow2024};
and MPC formulations that enforce a constant input over the entire horizon to reduce decision–variable dimension and computation time, typically with only minor performance degradation when horizon effects are small
%
%
\cite{makarow2018,taleb2018,dikarew2024}.

The function $f^{\ell_\rmh},$ in general, is nonlinear in $u$ and the construction of the safety filter is thus a complicated task. 
However, in the case of linear systems, the $\ell_{\rm h}$-step map $f^{\ell_{\rm h}}$ depends linearly on the input $u$ (under mild conditions detailed below).
Moreover, linearized models are routinely used for controller synthesis, as in standard linear MPC formulations
%
%
\cite{sanchez2017,igarashi2020,berberich2022}.



Consider the discrete-time system 
\begin{equation} \label{eq:linear_ss_dt}
    x_{k+1} = A x_k + B u_k.
\end{equation}
%
%
For a horizon $\ell_\rmh,$ it follows from \eqref{eq:linear_ss_dt} that the state at step $k + \ell_\rmh$ is given by
\begin{equation}\label{eq:linear_ss_dt_horizon}
    x_{k + \ell_\rmh} = A^{\ell_\rmh} x_k + \sum_{i = 0}^{\ell_\rmh  - 1} A^{i} B u_{k + i}.
\end{equation}
In the case where, at step $k,$ $u_{k + i} = u_k$ for all $i \in \{0, \ldots, \ell_\rmh  - 1\},$ it follows from \eqref{eq:linear_ss_dt_horizon} that the state at step $k + \ell_\rmh$ is given by
\begin{align}
    x_{k + \ell_\rmh} 
        &=
            A_{\ell_\rmh} x_k + B_{\ell_\rmh} u_k, \label{eq:linear_ss_dt_horizon_uk} 
\end{align} 
where
\begin{align}
    A_{\ell_\rmh} &\isdef A^{\ell_\rmh}, \quad 
    %
    B_{\ell_\rmh} \isdef \left(\sum_{i = 0}^{\ell_\rmh  - 1} A^{i} \right) B. \label{eq:B_pred}
\end{align}
%




    


The following proposition bounds the relative degree.
\begin{prop}
Consider the discrete-time system \eqref{eq:linear_ss_dt} and the corresponding system
\begin{align}
    \chi_{j+1} = A_{\ell_\rmh} \chi_j + B_{\ell_\rmh} v_j, 
    \label{eq:chi_dynamics}
\end{align}
where $j \isdef \ell_\rmh k$ and $v_j = u_{\ell_\rmh k}.$
If the relative degree of $h(x_k)$ with respect to $u_k$ is $\rho,$ then the relative degree of $h(\chi_j)$ with respect to $v_j$ is $\lceil\ell_\rmh/\rho \rceil$.    
\end{prop}
\textbf{Proof.}
    The relative degree $\rho$ of $(A,B,C)$ implies that $CA^{q}B=0$ for $q=0,\dots,\rho-2$ and $CA^{\rho-1}B\neq0$. 
    Consider the $\ell_h$-subsampled sequence with $j=\ell_h k$, $v_j=u_{\ell_h k}$. Then for the $m$-th sampled step,
    \vspace{-1em}
    \begin{align}        
        y_{j+m}
            &=
                y_{\ell_h k+m\ell_h}
            \nn \\
            &=
                CA^{m\ell_h}x_{\ell_h k}+\sum_{i=0}^{m\ell_h-1} CA^{m\ell_h-1-i}B\,u_{\ell_h k+i},
            \nn
    \end{align}
    and the coefficient multiplying $v_j=u_{\ell_h k}$ is the Markov parameter $CA^{m\ell_h-1}B$. 
    Note that $CA^{m\ell_h-1}B=0$ for $m\ell_h-1\le \rho-2$ and is nonzero at the smallest $m$ with $m\ell_h-1\ge \rho-1$, that is, $m\ell_h\ge \rho$. 
    Hence the relative degree of \eqref{eq:chi_dynamics}  is $r=\min\{m\in\mathbb{N}:m\ge \rho/\ell_h\}=\lceil \rho/\ell_h\rceil.$
    \hfill $\blacksquare$

It follows from the Proposition above that if the relative degree of $h(x_k)$ with respect to $u_k$ is $\ell_h,$ then the relative degree of $h(\chi_j)$ with respect to $v_j$ is 1.

%
%


Next, consider a polytopic control barrier function
\begin{equation}
    h(\chi_k) = A_{\rm cbf} \chi_k + b_{\rm cbf},
    \label{eq:poly_CBF}
\end{equation}
where $A_{\rm cbf} \in \BBR^{p \times n},$ $b_{\rm cbf} = \BBR^p$ and consider system \eqref{eq:chi_dynamics}.
The objective is to develop a safety filter such that $h(\chi_k) > 0.$
Note that the relative degree of $h(\chi_k)$ with respect to $v_k$ is one and thus the safety filter design process follows from Section \ref{sec:RD1}.
In particular, it follows from \eqref{eq:cbf_dt_v2} that the safety filter is
%
\begin{align}
    h\left(\chi_{k + 1}\right)  \ge \gamma  h(\chi_k). 
    \label{eq:mpcbf_1}
\end{align}
Substituting the polypotic control barrier function \eqref{eq:poly_CBF} in \eqref{eq:mpcbf_1} yields
\begin{align}
            A_{\rm cbf} A_{\ell_\rmh} x_k + A_{\rm cbf} B_{\ell_\rmh} u_k + b_{\rm cbf} 
        &\ge
            \gamma A_{\rm cbf} \chi_k + \gamma b_{\rm cbf}, \label{eq:mpcbf_2}
\end{align}
which simplifies to
\begin{align}
     \SA_\rms v_k 
        \ge
            \SB_\rms(\chi_k), \label{eq:mpcbf_3}
\end{align}
where
\begin{align}
    \SA_\rms & \isdef A_{\rm cbf} B_{\ell_\rmh} , 
    \\
    \SB_\rms(\chi_k)
        &\isdef 
            A_{\rm cbf} ( \gamma I_n -  A_{\ell_\rmh}) \chi_k - (1 - \gamma) b_{\rm cbf}.
\end{align}

\section{Numerical Simulations}\label{sec:simulations}

This section develops the safety filter based on the predictive CBF presented in the previous section and applies it to both a double integrator system with unknown input delay and the outer-loop controller of a bicopter lateral flight system.
The algorithm for the discrete-time LQR controller with integrator state and integrator anti-windup used for the nominal controllers in these examples is shown in Algorithm \ref{alg:LQR_Int}.
The implementation of the PCBF requires the solution of a constrained linear least-squares optimization problem. 
In all examples, this problem is solved by using the \texttt{lsqlin} solver from Matlab with the \texttt{active-set} algorithm.

\begin{algorithm}
    \caption{Discrete-time LQR controller with integrator state and integrator anti-windup, $\matl u_{{\rm nom}, k} & e_{{\rm int}, k} \matr^\rmT = {\rm LQR}_{\rm int}(e_k, e_{k-1}, u_{k-1}, e_{{\rm int}, k-1}, K_{\rm lqr}, \eta_{\rm aw}, C_{\rm int}, T_\rms)$}
    \label{alg:LQR_Int}
    \KwIn{Error state $e_k,$ previous error state $e_{k-1},$ previous control input $u_{k-1},$ previous integrator state $e_{{\rm int}, k-1},$ LQR gain $K_{\rm lqr},$ integrator anti-windup $\eta_{\rm aw},$ integrator state selection matrix $C_{\rm int},$ sampling time $T_\rms$}
    \KwOut{Nominal control input $u_{{\rm nom}, k},$ integrator state $e_{{\rm int}, k}$}
    $e_{{\rm dt}, k}$ $\leftarrow$ $(e_k - e_{k-1})/T_\rms$\\
    $e_{{\rm aug}, k}$ $\leftarrow$ $\matl e_{{\rm dt}, k} \\ C_{\rm int} e_k \matr$\\
    $u_{{\rm lqr}, k}$ $\leftarrow$ $K_{\rm lqr} \ e_{{\rm aug}, k}$ \\
    \nonl $\triangleright$ Calculate a preliminary LQR input.\\
    $u_{{\rm int}, k}$ $\leftarrow$ $u_{{\rm lqr}, k} + \eta_{\rm aw} (u_{k-1} - e_{{\rm int}, k-1})$ \\
    \nonl $\triangleright$ Calculate the integrator input by subtracting an anti-windup term.\\
    $e_{{\rm int}, k}$ $\leftarrow$ $e_{{\rm int}, k-1} + T_\rms u_{{\rm int}, k}$ \\
    \nonl $\triangleright$ Update the integrator state.\\
    $u_{{\rm nom}, k}$ $\leftarrow$ $e_{{\rm int}, k}$ 
\end{algorithm}

\subsection{Discretized double integrator}\label{subsec:ex_double_integrator}

Consider the discretized double integrator with an unknown input delay
\begin{align}
    x_{k+1} = A x_k + B u_{k-m}, \label{eq:dynamics_example_1}
\end{align}
where
\begin{align}
    A \isdef \matl 1 & T_\rms \\ 0 & 1 \matr, \quad 
    B \isdef \matl T_\rms^2/2 \\ T_\rms \matr,
    \label{eq:dynamics_example_1_AB}
\end{align}
$x_k = \matl x_{1, k} & x_{2, k} \matr^\rmT$ is the state, with $x_{1, k}, x_{2, k} \in \BBR,$ $u_k \in \BBR$ is the control input, 
and $m>0\in \BBZ$ is the unknown input delay. 
In this example, we set $T_\rms = 1.$
%
%
Let $r_k \in \BBR$ be a reference signal for $x_{1, k}.$ Hence, the objective of the controller is to minimize $\sum_{k = 0}^{\infty} \Vert r_k - x_{1, k}\Vert.$

Consider a nominal discrete-time LQR controller with an integrator state and an integrator anti-windup 
\begin{align}
    \matl u_{\rmr , k} \\ e_{{\rm int}, k} \matr = & {\rm LQR}_{\rm int} \left( \matl r_k \\ 0 \matr  - x_k, \matl r_{k-1} \\ 0 \matr - x_{k-1}, u_{k-1}, \right. \nn \\ &\qquad \qquad
    e_{{\rm int}, k-1}, K_{\rm lqr}, 0.2, \matl 1 & 0 \matr, T_\rms),
\end{align}
where $u_{\rmr, k} \in \BBR$ is the requested control input, the function ${\rm LQR}_{{\rm int}, k}$ is described in Algorithm \ref{alg:LQR_Int}, $e_{\rm int}$ is an internal integrator state, and $K_{\rm lqr}$ is the LQR gain obtained from solving the algebraic Ricatti equation with the state and input matrices shown in \eqref{eq:dynamics_example_1_AB} with additional dimensions to include an additional integrator state. 
Note that this controller is designed without accounting for the input delay $m.$
%

%
%

Next, consider the desired constraints 
\begin{equation*}
    x_{1, k} \in [x_{1, {\rm min}}, \ x_{1, {\rm max}}], \quad x_{2, k} \in [x_{2, {\rm min}}, \ x_{2, {\rm max}}],
\end{equation*}
where $x_{1, {\rm min}} < x_{1, {\rm max}} \in \BBR,$ and $x_{2, {\rm min}} < x_{2, {\rm max}} \in \BBR,$ which can be written in CPC form as
\begin{align}
    & h(x_k) = A_{\rm cbf} x_k + b_{\rm cbf} \ge 0, 
\end{align}
where
\begin{align}
    A_{\rm cbf} = & \matl 1 & 0 \\ -1 & 0 \\ 0 & 1 \\ 0 & -1 \matr, \quad b_{\rm cbf} = \matl -x_{1, {\rm min}} \\ x_{1, {\rm max}} \\ -x_{2, {\rm min}} \\ x_{2, {\rm max}} \matr.
\end{align}
%
%
The control input $u_k$ is obtained by solving the constrained linear least-squares optimization problem
\begin{equation}
    u_k = \argmin_{\nu \in \BBR} \lVert \nu - u_{\rmr, k} \rVert_2,
\end{equation}
subject to
\begin{equation}
    A_{\rm cbf} B_{\ell_\rmh} \ \nu \ge  A_{\rm cbf} (\gamma I_2 - A_{\ell_\rmh}) x_k - (1 - \gamma) b_{\rm cbf},
\end{equation}
where $\ell_\rmh > 0$ is the prediction horizon, $\gamma \in [0, 1],$ and $A_{\ell_\rmh}, B_{\ell_\rmh}$ are given by \eqref{eq:B_pred}, respectively, with $A$ and $B$ given by \eqref{eq:dynamics_example_1_AB}.

For all simulations, in the nominal controller, we set $x_0 = x_{-1} = u_{-1} = e_{{\rm int}, -1} = 0,$ and $K_{\rm lqr} = \matl  0.152 & 0.542 & 0.016\matr,$ which is obtained by the algorithm described in Algorithm \ref{alg:LQR_Int}. 
The unknown input delay is set to $m = 1.$
%

\begin{figure}[!ht]
    \vspace{-0.5em}
    \centering
    \includegraphics[width = \columnwidth]{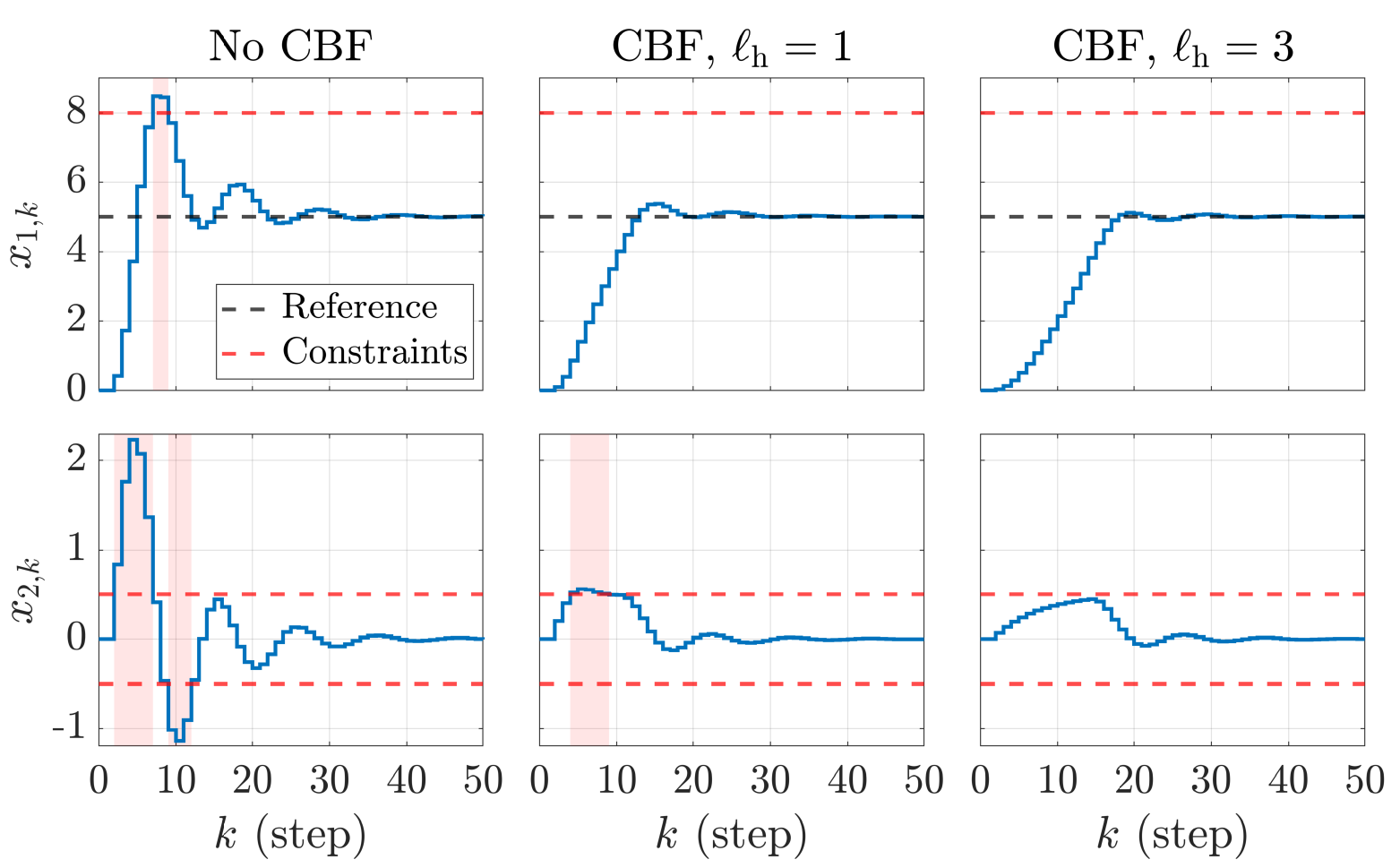}
    \caption{
    \textbf{Feasible command.}
    Closed-loop response of the discrete-time double integrator in the case of a feasible command  with a delay of $m = 1.$ 
    The command is given by $r_k \equiv 5,$ and the state constraints are defined by $x_{1,k} \in  [-8,  \ 8]$ and $x_{2, k} \in [-0.5, \ 0.5]$ for all $k \ge 0.$
    The responses in the cases with no CBF and with predictive CBF with $\gamma = 0.6, \ell_\rmh = 1$ and $\gamma = 0.6, \ell_\rmh = 3$ are shown.
    The red-shaded areas correspond to periods of time during which the constraints are violated.
    }
    \label{fig:ex_double_integrator_v1}
    \vspace{-1em}
\end{figure}

%
First, we consider the case of a feasible command.
The command is given by $r_k \equiv 5,$ and the state constraints are defined by $x_{1,k} \in  [-8,  \ 8]$ and $x_{2, k} \in [-0.5, \ 0.5]$ for all $k \ge 0,$ such that $x_{1,\rm max} = - x_{1,\rm min} = 8$ and $x_{2,\rm max} = - x_{2,\rm min} = 0.5.$
Figure \ref{fig:ex_double_integrator_v1} shows the closed-loop response with the safety filter for $\gamma = 0.6$ and two values of $\ell_\rmh.$
Note that the constraints are satisfied for all $k \ge 0$ when $\ell_\rmh>m.$

\begin{figure}[!ht]
    \vspace{1em}
    \centering
    \includegraphics[width = \columnwidth]{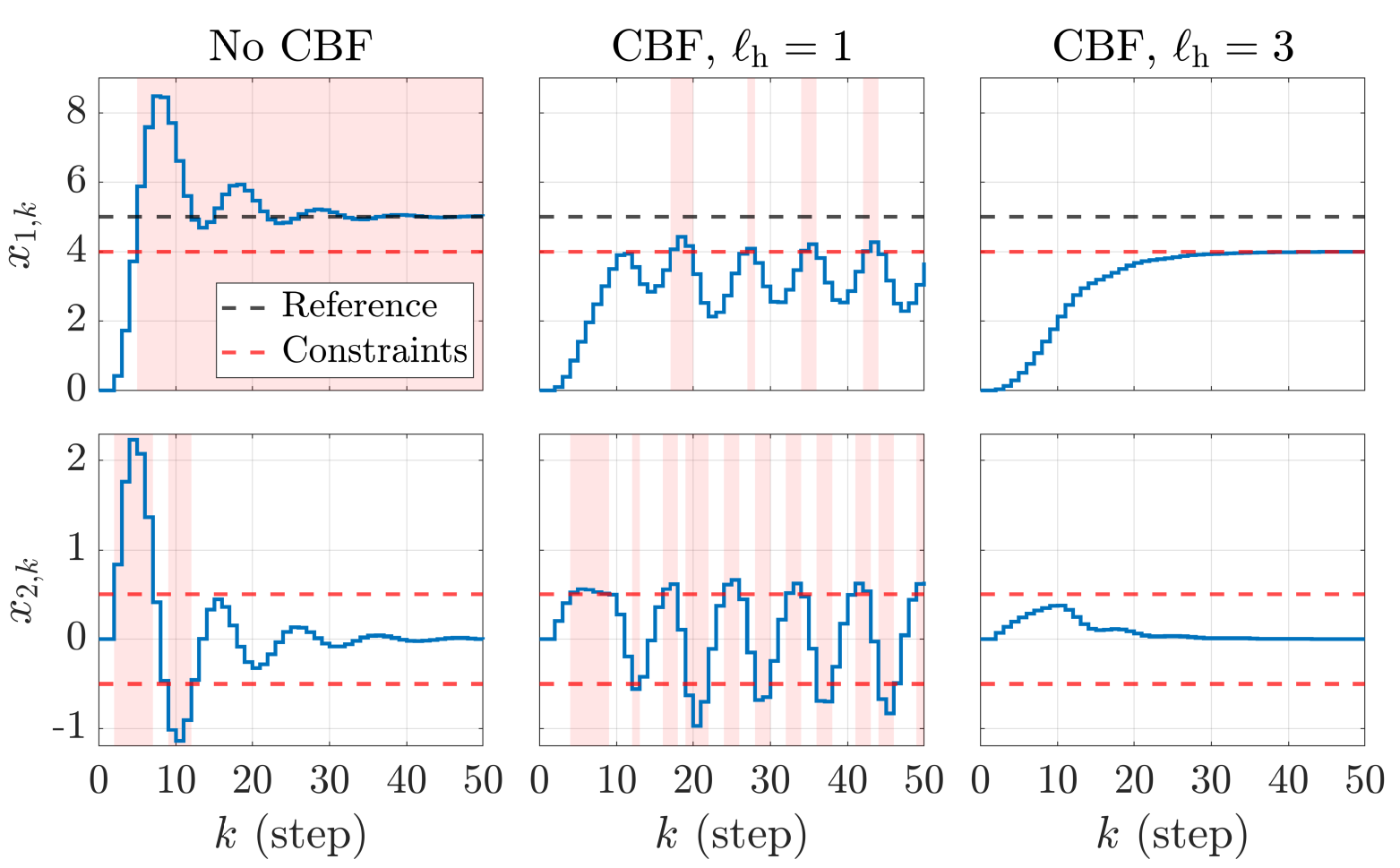}
    \caption{
    \textbf{Infeasible command.}
    Closed-loop response of the discrete-time double integrator in the case of an infeasible command with a delay of $m = 1.$ 
    The command is given by $r_k \equiv 5,$ and the state constraints are defined by $x_{1,k} \in  [-4,  \ 4]$ and $x_{2, k} \in [-0.5, \ 0.5]$ for all $k \ge 0.$
    The responses in the cases with no CBF and with predictive CBF with $\gamma = 0.6, \ell_\rmh = 1$ and $\gamma = 0.6, \ell_\rmh = 3$ are shown.
    The red-shaded areas correspond to periods of time during which the constraints are violated.
    }
    \label{fig:ex_double_integrator_v2}
    \vspace{-1em}
\end{figure}

Next, we consider the case of an infeasible command. 
The command $r_k \equiv 5$ and $x_{1,\rm max} = - x_{1,\rm min} = 4$ and $x_{2,\rm max} = - x_{2,\rm min} = 0.5.$
The command is given by $r_k \equiv 5,$ and the state constraints are defined by $x_{1,k} \in  [-4,  \ 4]$ and $x_{2, k} \in [-0.5, \ 0.5]$ for all $k \ge 0,$ such that $x_{1,\rm max} = - x_{1,\rm min} = 4$ and $x_{2,\rm max} = - x_{2,\rm min} = 0.5.$
Figure \ref{fig:ex_double_integrator_v2} shows the closed-loop response with the safety filter for $\gamma = 0.6$ and two values of $\ell_\rmh.$
Note that the constraints are satisfied for all $k \ge 0$ when $\ell_\rmh> m$.
Furthermore, note that the command is not followed in the cases in which CBF is applied since the command and the constraints cannot be simultaneously satisfied in this case.

\subsection{Bicopter lateral flight}\label{subsec:ex_multicopter}

\begin{figure}[!ht]
    \centering
    \includegraphics[width = 0.5\columnwidth]{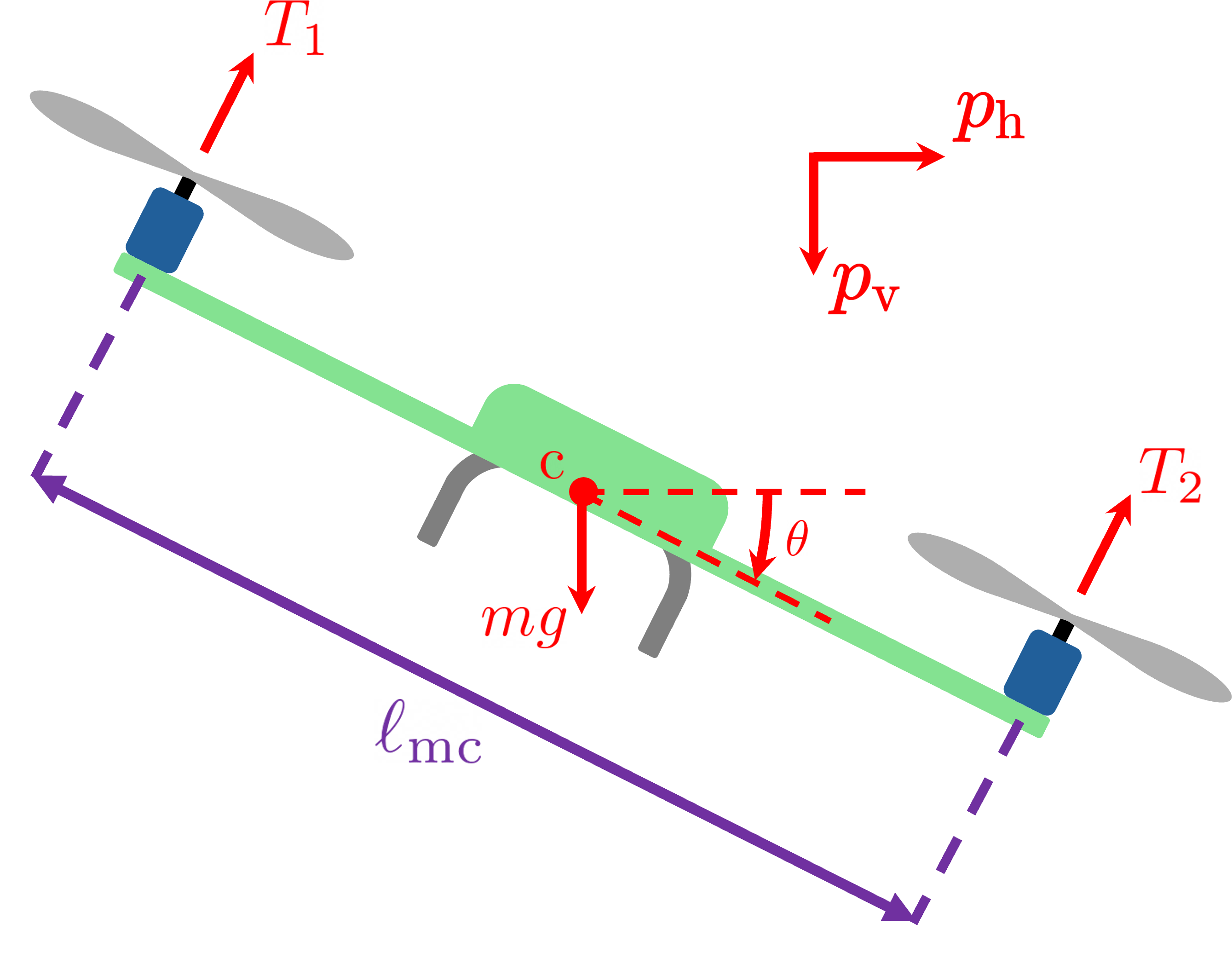}
    \caption{Diagram of bicopter in vertical plane}
    \label{fig:bicopter_diagram}
\end{figure}

Consider the bicopter in the vertical plane shown in Figure \ref{fig:bicopter_diagram}, which consists of a rigid frame with two rotors that generate thrust along their respective axes. 
The bicopter has mass $m$, center of mass $c$, moment of inertia $J$ about $c$, and the distance between the rotors is $\ell_{\rm mc}$. 
Let $T_1, T_2$ denote the thrusts produced by the left and right rotors, respectively, as shown in Figure \ref{fig:bicopter_diagram}. 
Define the total thrust $T \isdef T_1 + T_2$ and the total moment $\tau \isdef (T_1 - T_2)/\ell_{\rm mc}$. 
Then, the dynamics of the bicopter in the vertical plane are then given by
\small
\begin{align}
    \dot{p}_\rmh &= v_\rmh, &
    \dot{v}_\rmh &= \frac{T}{m} \sin \theta,  \label{eq:bicopter_dyn_1} \\ 
    \dot{p}_\rmv &= v_\rmv, &
    \dot{v}_\rmv &= -\frac{T}{m} \cos \theta + g, \\
    \dot{\theta} &= \omega, &
    \dot{\omega} &= \frac{\tau}{J}, \label{eq:bicopter_dyn_6}
\end{align}
\normalsize
where $p_\rmh, p_\rmv \in \BBR$ are the horizontal and vertical positions of $c,$ respectively, $v_\rmh, v_\rmv \in \BBR$ are the horizontal and vertical velocities of $c,$ respectively, $\theta \in \BBR$ is the bicopter tilt, $\omega \in \BBR$ is the bicopter angular velocity, and $g$ is the acceleration due to gravity.
In this example, let the state be given by $x \isdef \matl p_\rmh & v_\rmh & p_\rmv & v_\rmv & \theta & \omega. \matr^\rmT$

A discrete-time controller is implemented to control the continuous-time dynamics shown in \eqref{eq:bicopter_dyn_1}-\eqref{eq:bicopter_dyn_6}.
Hence, the states $p_\rmh, v_\rmh, p_\rmh, v_\rmh, \theta, \omega$ are sampled to produce the sampled states
\small
\begin{align*}
    p_{\rmh, k} &\isdef p_\rmh (k T_\rms), & v_{\rmh, k} &\isdef v_\rmh (k T_\rms), \\
    p_{\rmv, k} &\isdef p_\rmv (k T_\rms), & v_{\rmv, k} &\isdef v_\rmv (k T_\rms), \\
    \theta_k &\isdef \theta (k T_\rms), & \omega_k &\isdef \omega (k T_\rms),
\end{align*}
\normalsize
where $k \ge 0$ is the discrete-time step, and $T_\rms > 0$ is the sampling time.
The controller generates the total thrust $T_k \ge 0$ and the total torque $\tau_k \in \BBR.$
The continuous-time signals $T$ and $\tau$ applied to the bicopter are generated by applying a zero-order hold operation to $T_k$ and $\tau_k,$ that is, for all $k\ge0,$ 
\begin{equation}
    T(t) = T_k, \ \tau(t) = \tau_k, \mbox{ for all } t \in [kT_\rms, (k+1) T_\rms).
\end{equation}

\begin{figure} [h!]
    \vspace{-1em}
    \centering
    \resizebox{\columnwidth}{!}{%
    \begin{tikzpicture}[>={stealth'}, line width = 0.25mm]

    \node [input, name=ref]{};
    \node [sum, fill=green!20, right =0.75cm of ref] (sum2) {};
    \node[draw = white] at (sum2.center) {$+$};
    \node [smallblock, fill=green!20, rounded corners, right = 0.4cm of sum2 , minimum height = 0.6cm , minimum width = 0.7cm] (controller) {$G_{\rmc, {\rm ol}}$};
    \node [smallblock, fill=green!20, rounded corners, right = 0.75cm of controller.east , minimum height = 0.6cm , minimum width = 0.7cm] (cbf) {$f_{\rm cbf}$};
    \node [smallblock, fill=green!20, rounded corners, right = 0.75cm of cbf.east , minimum height = 1cm, minimum width = 0.7cm] (map) {$f_{\rm map}$};
    \node [smallblock, fill=green!20, rounded corners, below right = 1.25cm and 1.25 cm of map.center , minimum height = 0.6cm, minimum width = 0.7cm] (controller_IL) {$G_{\rmc, {\rm il}}$};
    \node [sum, fill=green!20, left = 0.3cm of controller_IL.west] (sum3) {};
    \node[draw = white] at (sum3.center) {$+$};
    \draw[->] ([xshift = -0.8cm]sum2.west) -- node [above, xshift = -0.2cm] {\scriptsize $\matl r_{\rmh, k} \\ 0 \\ r_{\rmv, k} \\ 0 \matr$} (sum2.west);
    \draw[->] ([yshift = -1.5cm]sum2.south) -- node [xshift = -0.5cm, yshift = -0.2cm] {\scriptsize $\matl p_{\rmh, k} \\ v_{\rmh, k} \\ p_{\rmv, k} \\ v_{\rmv, k} \matr$} node [xshift = 0.25 cm, yshift = 0.5cm] {$-$} (sum2.south);
    \draw[->] ([yshift = -1.3cm]sum2.south) -| (cbf.south);
    \draw[->] (sum2.east) -- (controller.west);
    \draw[->] (controller.east) -- node [above] {\small $u_{\rmr, k}$} (cbf.west);
    \draw[->] (cbf.east) -- node [above] {\small $u_k$} (map.west);
    \draw[->] ([yshift = -0.25cm]map.east) -| node [near end, xshift = 0.4cm, yshift = 0.1cm]{\scriptsize $\matl \theta_{\rmr, k} \\ 0 \matr$}(sum3.north);
    \draw[->] ([xshift = -2em]sum3.west) -- node [above, xshift = -0.25cm]{\scriptsize $\matl \theta_k \\ \omega_k \matr$} node [below, xshift = 0.15cm, yshift = 0.05cm]{$-$} (sum3.west);
    \draw[->] (sum3.east) -- (controller_IL.west);
    \draw[->] ([yshift = 0.25cm]map.east) -- node[above, near end, xshift = 0.2cm]{\small$T_k$} ([xshift = 2.5cm, yshift = 0.25cm]map.east);
    \draw[->] (controller_IL.east) -- ([xshift = 0.2cm]controller_IL.east) |- node[above, near end, xshift = -0.05cm]{\small$\tau_k$} ([xshift = 2.5cm, yshift = -0.25cm]map.east);
    \end{tikzpicture}
    }  
    \caption{Inner-loop, outer-loop control architecture considered for the bicopter example.}
    \label{bicopter_cbf_blk_diag}
    \vspace{-1em}
\end{figure}
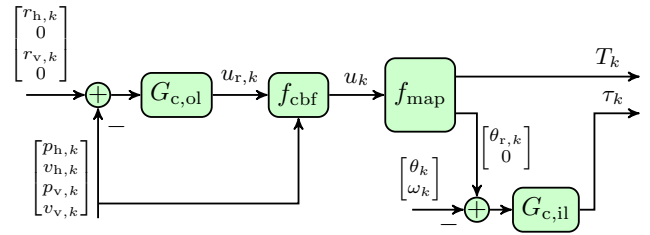

The controller is designed so that $p_{\rmh, k}$ and $p_{\rmv, k}$ follow reference signals $r_{\rmh, k}$ and $r_{\rmv, k},$ respectively, such that the objective of the controller is to minimize $\sum_{k = 0}^{\infty} \left\Vert \matl r_{\rmh, k} & r_{\rmv, k} \matr^\rmT - \matl p_{\rmh, k} & p_{\rmv, k} \matr^\rmT \right\Vert.$
For this purpose, the inner-loop, outer-loop control architecture shown in Figure \ref{bicopter_cbf_blk_diag} is adopted.
An advantage of this architecture is that it allows the dynamics shown in \eqref{eq:bicopter_dyn_1}-\eqref{eq:bicopter_dyn_6} to be decoupled into linear systems, such that the resulting decoupled, discretized dynamics are given by
\begin{align}
    x_{\rmh, k+1} & = A_{\rm pos} x_{\rmh, k} + B_{\rm pos} u_{\rmh, k}, \label{eq:bicopter_dsg_1}\\
    x_{\rmv, k+1} & = A_{\rm pos}x_{\rmv, k} + B_{\rm pos} u_{\rmv, k}, \\
    x_{{\rm att}, k+1} & = A_{\rm att} x_{{\rm att}, k} + B_{\rm att} \tau_k, \label{eq:bicopter_dsg_3}
\end{align}
where $x_{\rmh, k} \isdef \matl p_{\rmh, k} & v_{\rmh, k} \matr^\rmT,$ $x_{\rmv, k} \isdef \matl p_{\rmv, k} & v_{\rmv, k} \matr^\rmT,$ $x_{{\rm att}, k} \isdef \matl \theta_k & \omega_k \matr^\rmT,$ $u_{\rmh, k}, u_{\rmv, k} \in \BBR$ are horizontal and vertical acceleration commands, respectively, and 
\begin{align*}
    A_{\rm pos} \isdef A_{\rm att} &\isdef \matl 1 & T_\rms \\ 0 & 1 \matr, \\
    B_{\rm pos} \isdef \matl T_\rms^2 / (2 m) \\ T_\rms / m \matr, & \quad B_{\rm att} \isdef \matl T_\rms^2 / (2 J) \\ T_\rms / J \matr.
\end{align*}
The dynamics shown in \eqref{eq:bicopter_dsg_1}--\eqref{eq:bicopter_dsg_3} are used to design the outer-loop and inner-loop controllers

Let the outer-loop controller $G_{\rmc, {\rm ol}}$ be given by two LQR controllers for the horizontal and vertical states separately, such that
\small
\begin{align}
    \matl u_{\rmr, \rmh , k} \\ e_{{\rm int}, \rmh, k} \matr &= {\rm LQR}_{\rm int} \left( \matl r_{\rmh, k} \\ 0 \matr  - x_{\rmh, k}, \matl r_{\rmh, k-1} \\ 0 \matr - x_{\rmh, k-1}, \right. \nn \\ 
    &\qquad \qquad 
    u_{\rmh, k-1}, e_{{\rm int}, \rmh, k-1}, K_{{\rm lqr}, \rmh}, 0.2, \matl 1 & 0 \matr, T_\rms), \\
    \matl u_{\rmr, \rmv , k} \\ e_{{\rm int}, \rmv, k} \matr &= {\rm LQR}_{\rm int} \left( \matl r_{\rmv, k} \\ 0 \matr  - x_{\rmv, k}, \matl r_{\rmv, k-1} \\ 0 \matr - x_{\rmv, k-1}, \right. \nn \\ 
    &\qquad \qquad 
    u_{\rmv, k-1}, e_{{\rm int}, \rmv, k-1}, K_{{\rm lqr}, \rmv}, 0.2, \matl 1 & 0 \matr, T_\rms),
\end{align}
\normalsize
where the function ${\rm LQR}_{\rm int}$ is described in Algorithm \ref{alg:LQR_Int}, $e_{{\rm int}, \rmh, k}, e_{{\rm int}, \rmv, k}$ are the internal integrator states associated with the horizontal and vertical positions, respectively, and $K_{{\rm lqr}, \rmh},K_{{\rm lqr}, \rmv}$ are the LQR gains associated with the horizontal and vertical positions, respectively, obtained from solving the algebraic Ricatti equation with the state matrix $A_{\rm pos}$ and input matrix $B_{\rm pos}$ with additional dimensions to include an additional integrator state. 
Note that $K_{{\rm lqr}, \rmh}$ and $K_{{\rm lqr}, \rmv}$ are designed separately.

Next, in this example, the function $f_{\rm cbf}$ implements PCBF for translational position and velocity constraint enforcement,
whose inputs $u_{\rmr, k} \isdef \matl u_{\rmr, \rmh , k} & u_{\rmr, \rmv , k} \matr^\rmT$ and $\matl x_{\rmh, k} & x_{\rmv, k} \matr^\rmT,$ and its output is $u_k \isdef \matl u_{\rmh , k} & u_{\rmv , k} \matr^\rmT.$
More PCBF implementation details are given later in \eqref{eq:bicopter_cbf_f_1}--\eqref{eq:bicopter_cbf_f_3}.

The outer and inner loops are linked by a nonlinear mapping function $f_{\rm map}$ that can be used to obtain the thrust $T_k$ and a reference tilt value $\theta_{\rmr, k}$ from $u_k,$ such that
\begin{equation}
    \matl T_k \\ \theta_{\rmr, k} \matr = f_{\rm map} (u_k) = \matl \sqrt{ u_{\rmh, k}^2 + \left(mg -u_{\rmv, k}\right)^2 } \\ {\rm atan2}(u_{\rmh, k}, \ mg - u_{\rmv, k}) \matr.
\end{equation}
Then, the inner-loop controller $G_{\rmc, {\rm il}}$ is given by a LQR controller for the attitude states, such that
\small
\begin{align}
    \matl \tau_k \\ e_{{\rm int}, {\rm att}, k} \matr &= {\rm LQR}_{\rm int} \left( \matl \theta_{\rmr, k} \\ 0 \matr  - x_{{\rm att}, k}, \matl \theta_{\rmr, k-1} \\ 0 \matr  - x_{{\rm att}, k-1}, \right. \nn \\ 
    &\qquad \qquad 
    \tau_{k-1}, e_{{\rm int}, {\rm att}, k-1}, K_{{\rm lqr}, {\rm att}}, 0.2, \matl 1 & 0 \matr, T_\rms),
\end{align}
\normalsize
where the function ${\rm LQR}_{\rm int}$ is described in Algorithm \ref{alg:LQR_Int}, $e_{{\rm int}, {\rm att}, k}$ is the internal integrator state associated with the tilt, and $K_{{\rm lqr}, {\rm att}}$ is the LQR gain associated with the attitude states, obtained from solving the algebraic Ricatti equation with the state matrix $A_{\rm att}$ and input matrix $B_{\rm att}$ with additional dimensions to include an additional integrator state.

Next, consider the desired constraints 
\begin{align*}
    p_{\rmh, k} &\in [p_{\rmh, {\rm min}}, p_{\rmh, {\rm max}}], & v_{\rmh , k} &\in [v_{\rmh , {\rm min}}, \ v_{\rmh , {\rm max}}], \\
    p_{\rmv, k} &\in [p_{\rmv, {\rm min}}, p_{\rmv, {\rm max}}], & v_{\rmv , k} &\in [v_{\rmv , {\rm min}}, \ v_{\rmv , {\rm max}}],
\end{align*}
where $p_{\rmh, {\rm min}} < p_{\rmh, {\rm max}} \in \BBR,$ $v_{\rmh , {\rm min}} <v_{\rmh , {\rm max}} \in \BBR,$ $p_{\rmv, {\rm min}} < p_{\rmv, {\rm max}} \in \BBR,$ and $v_{\rmv , {\rm min}} <v_{\rmv , {\rm max}} \in \BBR,$.
These constraints can be written in CPC form as
\begin{align}
    h_\rmh (x_{\rmh, k}) &= A_{\rm cbf} x_{\rmh, k} + b_{{\rm cbf}, \rmh} \ge 0, \\
    h_\rmv (x_{\rmv, k}) &= A_{\rm cbf} x_{\rmv, k} + b_{{\rm cbf}, \rmv} \ge 0, 
\end{align}
where
\footnotesize
\begin{align}
    A_{\rm cbf} = & \matl 1 & 0 \\ -1 & 0 \\ 0 & 1 \\ 0 & -1 \matr, \ b_{{\rm cbf}, \rmh} = \matl -p_{\rmh, {\rm min}} \\ p_{\rmh, {\rm max}} \\ -v_{\rmh ,  {\rm min}} \\ v_{\rmh , {\rm max}} \matr, \ b_{{\rm cbf}, \rmv} = \matl -p_{\rmv, {\rm min}} \\ p_{\rmv, {\rm max}} \\ -v_{\rmv ,  {\rm min}} \\ v_{\rmv , {\rm max}}  \matr.
\end{align}
\normalsize
Hence, the control input $u_k$ is obtained by solving the constrained linear least-squares optimization problem
\begin{equation}
    u_k = \argmin_{ [\nu_1 \ \nu_2 ]^\rmT \in \BBR^2} \left\lVert \matl \nu_1 \\ \nu_2 \matr - u_{\rmr, k} \right\rVert_2, \label{eq:bicopter_cbf_f_1}
\end{equation}
subject to
\small
\begin{align}
    A_{\rm cbf} B_{\ell_{\rmh, \rmh}} \ \nu_1 \ge  A_{\rm cbf} (\gamma I_2 - A_{\ell_{\rmh, \rmh}}) x_{\rmh, k} - (1 - \gamma) b_{\rm cbf},\\
    A_{\rm cbf} B_{\ell_{\rmh, \rmv}} \ \nu_2 \ge  A_{\rm cbf} (\gamma I_2 - A_{\ell_{\rmh, \rmv}}) x_{\rmv, k} - (1 - \gamma) b_{\rm cbf},  \label{eq:bicopter_cbf_f_3}
\end{align}
\normalsize
where $\ell_{\rmh, \rmh}, \ell_{\rmh, \rmv} > 0$ are the prediction horizons associated with horizontal and vertical dynamics, respectively, $\gamma \in [0, 1],$ 
$A_{\ell_{\rmh, \rmh}}, B_{\ell_{\rmh, \rmh}}$ are given by \eqref{eq:B_pred}, respectively, with $A = A_{\rm pos},$ $B = B_{\rm pos},$ and $\ell_\rmh = \ell_{\rmh, \rmh},$
and 
$A_{\ell_{\rmh, \rmv}}, B_{\ell_{\rmh, \rmv}}$ are given by \eqref{eq:B_pred}, respectively, with $A = A_{\rm pos},$ $B = B_{\rm pos},$ and $\ell_\rmh = \ell_{\rmh, \rmv}.$
Note that the design of PCBF does not account for the inner-loop controller $G_{\rmc, {\rm il}}$ and the tilt state dynamics \eqref{eq:bicopter_dyn_6}, which introduce unmodeled dynamics, and thus, an unknown relative degree to the problem formulation, whose effect is later shown to be mitigated by PCBF.

\begin{figure}[!ht]
    \centering
    \includegraphics[width = \columnwidth]{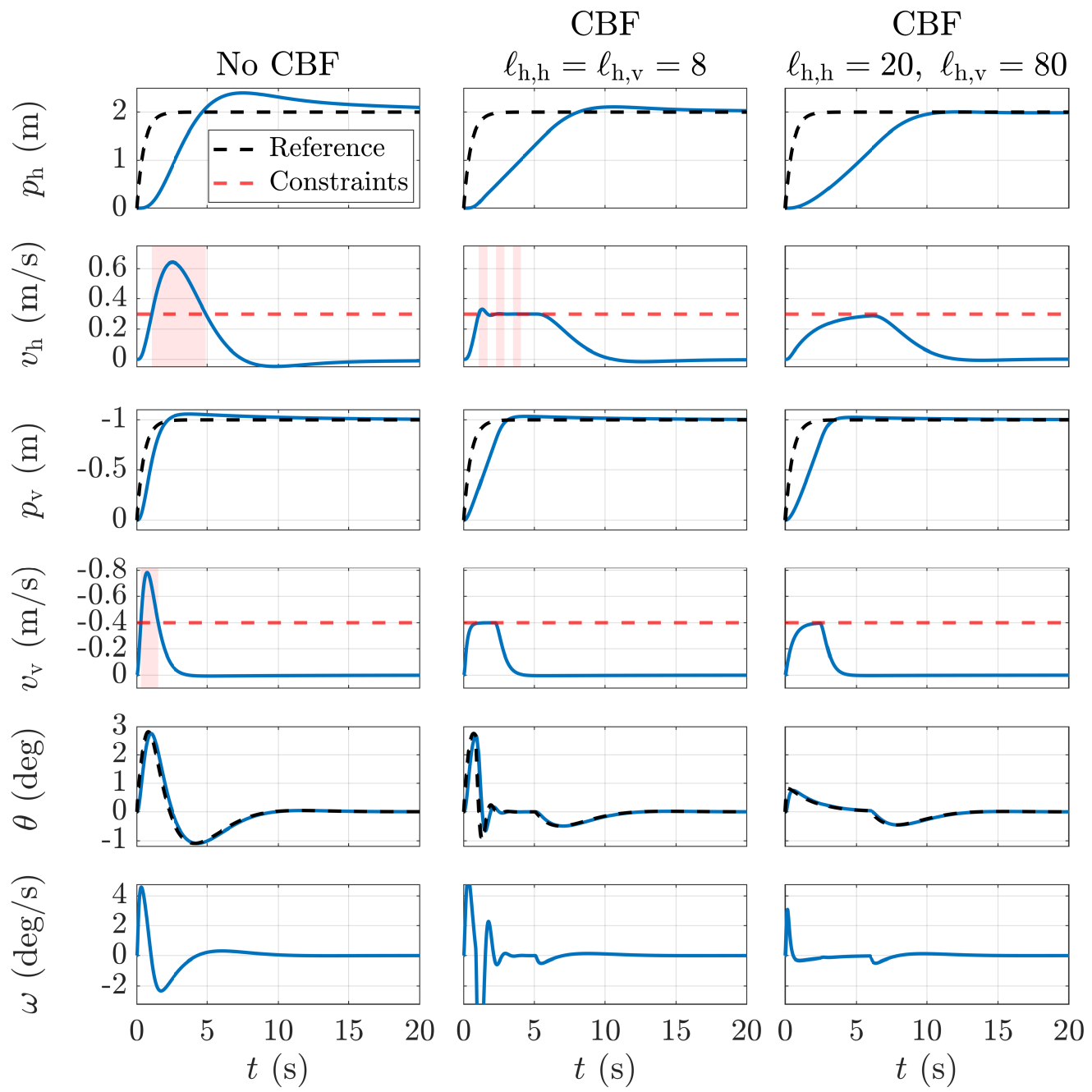}
    \caption{
    \textbf{Feasible command.}
    Closed-loop response of the bicopter in the vertical plane in the case of a feasible command.
    The command is shown as a dashed, black plot, and state constraints are defined by $p_\rmh \in [-3, \ 3]$ m, $v_\rmh \in [-0.3, \ 0.3]$ m/s, $p_\rmv \in [-3, \ 3]$ m, and $v_\rmv \in [-0.4, \ 0.4]$ m/s for all $t \ge 0.$
    The results in the cases with no CBF and with predictive CBF with $\gamma = 0.8, \ell_{\rmh, \rmh} = \ell_{\rmh, \rmv} = 8$ and $\gamma = 0.8,  \ell_{\rmh, \rmh} = 20, \ell_{\rmh, \rmv} = 80$ are shown.
    The red shaded areas correspond to periods of time in which the constraints are violated.
    }
    \label{fig:pcbf_ex_2_v1}
    \vspace{-1.5em}
\end{figure}

For all simulations, $x (t) = 0, T (t) = mg, \tau (t) = 0$ for all $t \le 0,$ 
$u_{\rmh, k} = u_{\rmv, k} = r_{\rmh, k} = r_{\rmv, k} = \theta_{\rmr, k} = e_{{\rm int}, \rmh, k} = e_{{\rm int}, \rmv, k} = e_{{\rm int}, {\rm att}, k} = 0$ for all $k \le 0,$
\footnotesize
\begin{equation*}
    K_{{\rm lqr}, \rmh} = \matl 0.397 \\ 0.918 \\ 0.032 \matr^\rmT, \ K_{{\rm lqr}, \rmv} = \matl 10.705 \\ 6.3849 \\ 1.392 \matr^\rmT, \ K_{{\rm lqr}, {\rm att}} = \matl 21.307 \\ 4.182 \\ 0.670 \matr^\rmT,
\end{equation*}
\normalsize
and $T_{\rm s} = 0.005$ s.
The Simulink environment is used for numerical simulation with the \texttt{ode45} solver to solve the bicopter continuous-time dynamics. 
The discrete-time dynamics corresponding to the controller and PCBF are evaluated every $T_{\rm s}$ seconds.
First, we consider the case of a feasible command.
The commands are shown in Figure \ref{fig:pcbf_ex_2_v1}, such that $\lim_{k \to \infty} r_{\rmh, k} = 2, \lim_{k \to \infty} r_{\rmv, k} = -1,$ and the state constraints are defined by $p_\rmh \in [-3, \ 3]$ m, $v_\rmh \in [-0.3, \ 0.3]$ m/s, $p_\rmv \in [-3, \ 3]$ m, and $v_\rmv \in [-0.4, \ 0.4]$ m/s for all $t \ge 0.$
Figure \ref{fig:pcbf_ex_2_v1} shows the closed-loop response with the safety filters for $\gamma = 0.8$ and two sets of values of $\ell_{\rmh, \rmh}, \ell_{\rmh, \rmv}.$
Note that the constraints are satisfied for all $k \ge 0$ in the case with the larger prediction horizons.

\begin{figure}[!ht]
    \centering
    \includegraphics[width = \columnwidth]{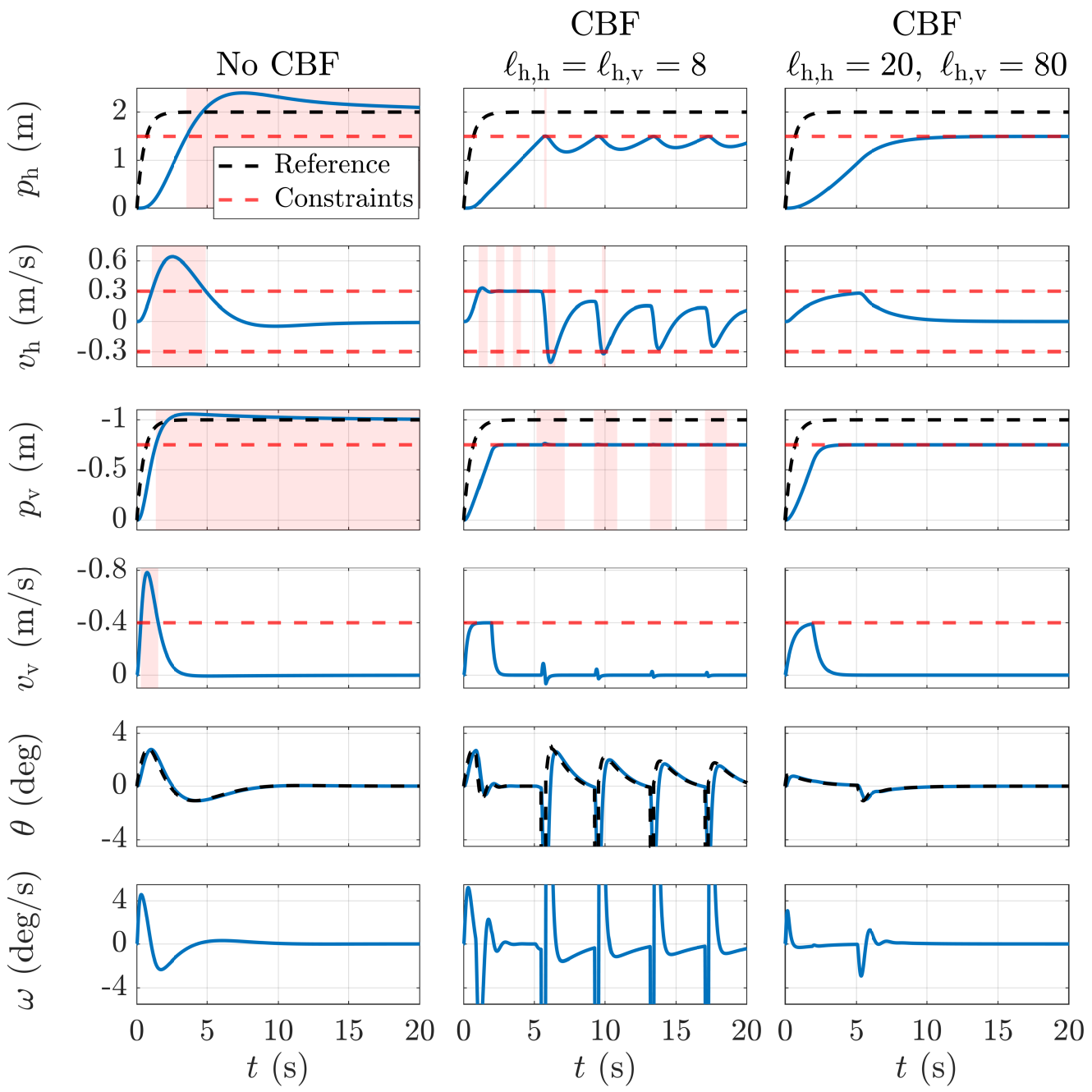}
    \caption{
    \textbf{Unfeasible command.}
    Closed-loop response of the bicopter in the vertical plane in the case of an unfeasible command.
    The command is shown as a dashed, black plot, and state constraints are defined by $p_\rmh \in [-1.5, \ 1.5]$ m, $v_\rmh \in [-0.3, \ 0.3]$ m/s, $p_\rmv \in [-0.75, \ 0.75]$ m, and $v_\rmv \in [-0.4, \ 0.4]$ m/s for all $t \ge 0.$
    The results in the cases with no CBF and with predictive CBF with $\gamma = 0.8, \ell_{\rmh, \rmh} = \ell_{\rmh, \rmv} = 8$ and $\gamma = 0.8, \ell_{\rmh, \rmh} = 20, \ell_{\rmh, \rmv} = 80$ are shown.
    The red shaded areas correspond to periods of time in which the constraints are violated.
    }
    \label{fig:pcbf_ex_2_v2}
    \vspace{-1em}
\end{figure}

Next, we consider the case of an infeasible command. 
The commands are shown in Figure \ref{fig:pcbf_ex_2_v2}, such that $\lim_{k \to \infty} r_{\rmh, k} = 2, \lim_{k \to \infty} r_{\rmv, k} = -1,$ and the state constraints are defined by $p_\rmh \in [-1.5, \ 1.5]$ m, $v_\rmh \in [-0.3, \ 0.3]$ m/s, $p_\rmv \in [-0.75, \ 0.75]$ m, and $v_\rmv \in [-0.4, \ 0.4]$ m/s for all $t \ge 0.$
Figure \ref{fig:pcbf_ex_2_v2} shows the closed-loop response with the safety filters for $\gamma = 0.8$ and two sets of values of $\ell_{\rmh, \rmh}, \ell_{\rmh, \rmv}.$
Note that the constraints are satisfied for all $k \ge 0$ in the case with the larger prediction horizons.
Furthermore, note that the command is not followed in the cases in which CBF is applied since the command and the constraints cannot be simultaneously satisfied in this case. 
A video showing an animation of the results in Figure \ref{fig:pcbf_ex_2_v2} is available at \href{https://youtu.be/G8rpmjDqxg0}{https://youtu.be/G8rpmjDqxg0}.
Furthermore, the results from another case with an unfeasible command and an octagonal potition constraint are shown \href{https://youtu.be/4y087R8wWaE}{https://youtu.be/4y087R8wWaE}, with $\gamma = 0.95$ and $\ell_{\rmh, \rmh} = \ell_{\rmh, \rmv} = 20.$

\section{Conclusions}
\label{sec:conclusions}

This paper introduced a predictive control barrier function (PCBF) formulation for enforcing state constraints in discrete-time systems with unknown relative degree caused by input delays or unmodeled input dynamics.
The effectiveness of the proposed technique was demonstrated in numerical simulations, including a discrete-time, double integrator with unknown input delay and a bicopter with position and velocity constraints, in which PCBF is implemented only after the outer-loop controller, which results in the introduction of unmodeled input dynamics composed by the inner-loop controller and the attitude dynamics.
Future work aims to implement this in a laboratory experiment and extend this formulation to be applicable to systems with nonconvex constraints and to include piecewise-linear models, as demonstrated in \cite{islam2025}, as well as piecewise-affine models.

\bibliographystyle{IEEEtran}
\bibliography{IEEEabrv,biblio}

\begin{thebibliography}{10}
\providecommand{\url}[1]{#1}
\csname url@samestyle\endcsname
\providecommand{\newblock}{\relax}
\providecommand{\bibinfo}[2]{#2}
\providecommand{\BIBentrySTDinterwordspacing}{\spaceskip=0pt\relax}
\providecommand{\BIBentryALTinterwordstretchfactor}{4}
\providecommand{\BIBentryALTinterwordspacing}{\spaceskip=\fontdimen2\font plus
\BIBentryALTinterwordstretchfactor\fontdimen3\font minus
  \fontdimen4\font\relax}
\providecommand{\BIBforeignlanguage}[2]{{%
\expandafter\ifx\csname l@#1\endcsname\relax
\typeout{** WARNING: IEEEtran.bst: No hyphenation pattern has been}%
\typeout{** loaded for the language `#1'. Using the pattern for}%
\typeout{** the default language instead.}%
\else
\language=\csname l@#1\endcsname
\fi
#2}}
\providecommand{\BIBdecl}{\relax}
\BIBdecl

\bibitem{menner2024translation}
M.~Menner and E.~Lavretsky, ``Translation of nagumo's foundational work on
  barrier functions: On the location of integral curves of ordinary
  differential equations,'' \emph{arXiv preprint arXiv:2406.18614}, 2024.

\bibitem{ames2019}
A.~D. Ames, S.~Coogan, M.~Egerstedt, G.~Notomista, K.~Sreenath, and P.~Tabuada,
  ``Control barrier functions: {Theory} and applications,'' in \emph{Proc.
  Europ. Contr. Conf.}\hskip 1em plus 0.5em minus 0.4em\relax IEEE, 2019, pp.
  3420--3431.

\bibitem{tan2021}
X.~Tan, W.~S. Cortez, and D.~V. Dimarogonas, ``High-order barrier functions:
  Robustness, safety, and performance-critical control,'' \emph{IEEE Trans.
  Automat. Contr.}, vol.~67, no.~6, pp. 3021--3028, 2021.

\bibitem{garg2024}
K.~Garg, J.~Usevitch, J.~Breeden, M.~Black, D.~Agrawal, H.~Parwana, and
  D.~Panagou, ``Advances in the theory of control barrier functions:
  {Addressing} practical challenges in safe control synthesis for autonomous
  and robotic systems,'' \emph{Ann. Rev. Contr.}, vol.~57, p. 100945, 2024.

\bibitem{breeden2021}
J.~Breeden, K.~Garg, and D.~Panagou, ``Control barrier functions in
  sampled-data systems,'' \emph{IEEE Contr. Sys. Lett.}, vol.~6, pp. 367--372,
  2021.

\bibitem{usevitch2022}
J.~Usevitch and D.~Panagou, ``Adversarial resilience for sampled-data systems
  under high-relative-degree safety constraints,'' \emph{IEEE Trans. Automat.
  Contr.}, vol.~68, no.~3, pp. 1537--1552, 2022.

\bibitem{tan2025}
X.~Tan, E.~Da{\c{s}}, A.~D. Ames, and J.~W. Burdick, ``Zero-order control
  barrier functions for sampled-data systems with state and input dependent
  safety constraints,'' in \emph{Proc. Amer. Contr. Conf.}\hskip 1em plus 0.5em
  minus 0.4em\relax IEEE, 2025, pp. 283--290.

\bibitem{jankovic2018}
M.~Jankovic, ``Robust control barrier functions for constrained stabilization
  of nonlinear systems,'' \emph{Automatica}, vol.~96, pp. 359--367, 2018.

\bibitem{zhang2022}
Y.~Zhang, S.~Walters, and X.~Xu, ``Control barrier function meets interval
  analysis: {Safety}-critical control with measurement and actuation
  uncertainties,'' in \emph{Proc. Amer. Contr. Conf.}\hskip 1em plus 0.5em
  minus 0.4em\relax IEEE, 2022, pp. 3814--3819.

\bibitem{bahati2024}
G.~Bahati, P.~Ong, and A.~D. Ames, ``Sample-and-hold safety with control
  barrier functions,'' in \emph{Proc. Amer. Contr. Conf.}\hskip 1em plus 0.5em
  minus 0.4em\relax IEEE, 2024, pp. 5169--5176.

\bibitem{alan2025}
A.~Alan, T.~G. Molnar, A.~D. Ames, and G.~Orosz, ``Generalizing robust control
  barrier functions from a controller design perspective,'' \emph{IEEE Open J.
  Contr. Syst.}, 2025.

\bibitem{yang2019}
G.~Yang, C.~Belta, and R.~Tron, ``Self-triggered control for safety critical
  systems using control barrier functions,'' in \emph{Proc. Amer. Contr.
  Conf.}\hskip 1em plus 0.5em minus 0.4em\relax IEEE, 2019, pp. 4454--4459.

\bibitem{taylor2020}
A.~J. Taylor, P.~Ong, J.~Cort{\'e}s, and A.~D. Ames, ``Safety-critical event
  triggered control via input-to-state safe barrier functions,'' \emph{IEEE
  Contr. Sys. Lett.}, vol.~5, no.~3, pp. 749--754, 2020.

\bibitem{sabouni2024}
E.~Sabouni, C.~G. Cassandras, W.~Xiao, and N.~Meskin, ``Optimal control of
  connected automated vehicles with event/self-triggered control barrier
  functions,'' \emph{Automatica}, vol. 162, p. 111530, 2024.

\bibitem{agrawal2017}
A.~Agrawal and K.~Sreenath, ``Discrete control barrier functions for
  safety-critical control of discrete systems with application to bipedal robot
  navigation.'' in \emph{Robotics: Science and Systems}, vol.~13, 2017, pp.
  1--10.

\bibitem{xiong2022}
Y.~Xiong, D.-H. Zhai, M.~Tavakoli, and Y.~Xia, ``Discrete-time control barrier
  function: High-order case and adaptive case,'' \emph{IEEE Trans. Cybernet.},
  vol.~53, no.~5, pp. 3231--3239, 2022.

\bibitem{zheng2024}
J.~Zheng, J.~Miller, and M.~Sznaier, ``Data-driven safe control of
  discrete-time non-linear systems,'' \emph{IEEE Contr. Sys. Lett.}, vol.~8,
  pp. 1553--1558, 2024.

\bibitem{ma2021}
H.~Ma, X.~Zhang, S.~E. Li, Z.~Lin, Y.~Lyu, and S.~Zheng, ``Feasibility
  enhancement of constrained receding horizon control using generalized control
  barrier function,'' in \emph{Proc. Int. Conf. Indust. Cyber-Phys.
  Syst.}\hskip 1em plus 0.5em minus 0.4em\relax IEEE, 2021, pp. 551--557.

\bibitem{zeng2021_2}
J.~Zeng, Z.~Li, and K.~Sreenath, ``Enhancing feasibility and safety of
  nonlinear model predictive control with discrete-time control barrier
  functions,'' in \emph{Proc. Conf. Dec. Contr.}\hskip 1em plus 0.5em minus
  0.4em\relax IEEE, 2021, pp. 6137--6144.

\bibitem{wabersich2022}
K.~P. Wabersich and M.~N. Zeilinger, ``Predictive control barrier functions:
  Enhanced safety mechanisms for learning-based control,'' \emph{IEEE Trans.
  Automat. Contr.}, vol.~68, no.~5, pp. 2638--2651, 2022.

\bibitem{liu2023_MPC_CBF}
S.~Liu, J.~Zeng, K.~Sreenath, and C.~A. Belta, ``Iterative convex optimization
  for model predictive control with discrete-time high-order control barrier
  functions,'' in \emph{Proc. Amer. Contr. Conf.}\hskip 1em plus 0.5em minus
  0.4em\relax IEEE, 2023, pp. 3368--3375.

\bibitem{hall2023}
A.~W. Hall, M.~Greeff, and A.~P. Schoellig, ``Differentially flat
  learning-based model predictive control using a stability, state, and input
  constraining safety filter,'' \emph{IEEE Contr. Sys. Lett.}, vol.~7, pp.
  2191--2196, 2023.

\bibitem{liu2025}
S.~Liu, Y.~Mao, and C.~A. Belta, ``Safety-critical planning and control for
  dynamic obstacle avoidance using control barrier functions,'' in \emph{Proc.
  Amer. Contr. Conf.}\hskip 1em plus 0.5em minus 0.4em\relax IEEE, 2025, pp.
  348--354.

\bibitem{priess2025}
M.~C. Priess, ``A convex model predictive control barrier function approach for
  differentially flat systems,'' \emph{IEEE Trans. Contr. Syst. Tech.}, 2025.

\bibitem{jankovic2018_delay}
M.~Jankovic, ``Control barrier functions for constrained control of linear
  systems with input delay,'' in \emph{Proc. Amer. Contr. Conf.}\hskip 1em plus
  0.5em minus 0.4em\relax IEEE, 2018, pp. 3316--3321.

\bibitem{kiss2023}
A.~K. Kiss, T.~G. Molnar, A.~D. Ames, and G.~Orosz, ``Control barrier
  functionals: {Safety}-critical control for time delay systems,'' \emph{Int.
  J. Rob. Nonlin. Contr.}, vol.~33, no.~12, pp. 7282--7309, 2023.

\bibitem{seiler2021}
P.~Seiler, M.~Jankovic, and E.~Hellstrom, ``Control barrier functions with
  unmodeled input dynamics using integral quadratic constraints,'' \emph{IEEE
  Contr. Syst. Lett.}, vol.~6, pp. 1664--1669, 2021.

\bibitem{quan2023}
Y.~S. Quan, J.~S. Kim, S.-H. Lee, and C.~C. Chung, ``Tube-based control barrier
  function with integral quadratic constraints for unknown input delay,''
  \emph{IEEE Contr. Syst. Lett.}, vol.~7, pp. 1730--1735, 2023.

\bibitem{kolmanovsky2014}
I.~Kolmanovsky, E.~Garone, and S.~Di~Cairano, ``Reference and command
  governors: {A} tutorial on their theory and automotive applications,'' in
  \emph{Proc. Amer. Contr. Conf.}\hskip 1em plus 0.5em minus 0.4em\relax IEEE,
  2014, pp. 226--241.

\bibitem{garone2017}
E.~Garone, S.~Di~Cairano, and I.~Kolmanovsky, ``Reference and command governors
  for systems with constraints: {A} survey on theory and applications,''
  \emph{Automatica}, vol.~75, pp. 306--328, 2017.

\bibitem{garone2018}
E.~Garone, M.~Nicotra, and L.~Ntogramatzidis, ``Explicit reference governor for
  linear systems,'' \emph{Int. J. Contr.}, vol.~91, no.~6, pp. 1415--1430,
  2018.

\bibitem{cagienard2007}
R.~Cagienard, P.~Grieder, E.~C. Kerrigan, and M.~Morari, ``Move blocking
  strategies in receding horizon control,'' \emph{J. Proc. Contr.}, vol.~17,
  no.~6, pp. 563--570, 2007.

\bibitem{shekhar2015}
R.~C. Shekhar and C.~Manzie, ``Optimal move blocking strategies for model
  predictive control,'' \emph{Automatica}, vol.~61, pp. 27--34, 2015.

\bibitem{makarow2024}
A.~Makarow, C.~R{\"o}smann, and T.~Bertram, ``Suboptimal nonlinear model
  predictive control with input move-blocking,'' \emph{Int. J. Contr.},
  vol.~97, no.~3, pp. 450--459, 2024.

\bibitem{makarow2018}
A.~Makarow, M.~Keller, C.~R{\"o}smann, and T.~Bertram, ``Model predictive
  trajectory set control with adaptive input domain discretization,'' in
  \emph{Proc. Amer. Contr. Conf.}\hskip 1em plus 0.5em minus 0.4em\relax IEEE,
  2018, pp. 3159--3164.

\bibitem{taleb2018}
M.~Taleb, E.~Leclercq, and D.~Lefebvre, ``Model predictive control for discrete
  and continuous timed petri nets,'' \emph{Int. J. Automat. Comput.}, vol.~15,
  no.~1, pp. 25--38, 2018.

\bibitem{dikarew2024}
A.~Dikarew, ``Helicopter short-term collision avoidance with sampling-based
  model predictive control,'' \emph{CEAS Aeronaut. J.}, vol.~15, no.~2, pp.
  385--396, 2024.

\bibitem{sanchez2017}
G.~S{\'a}nchez, M.~Murillo, L.~Genzelis, N.~Deniz, and L.~Giovanini, ``{MPC}
  for nonlinear systems: {A} comparative review of discretization methods,'' in
  \emph{Proc. Workshop Inf. Process. Contr.}\hskip 1em plus 0.5em minus
  0.4em\relax IEEE, 2017, pp. 1--6.

\bibitem{igarashi2020}
Y.~Igarashi, M.~Yamakita, J.~Ng, and H.~H. Asada, ``{MPC} performances for
  nonlinear systems using several linearization models,'' in \emph{Proc. Amer.
  Contr. Conf.}\hskip 1em plus 0.5em minus 0.4em\relax IEEE, 2020, pp.
  2426--2431.

\bibitem{berberich2022}
J.~Berberich, J.~K{\"o}hler, M.~A. M{\"u}ller, and F.~Allg{\"o}wer, ``Linear
  tracking {MPC} for nonlinear systems—{Part I}: {The} model-based case,''
  \emph{IEEE Trans. Automat. Contr.}, vol.~67, no.~9, pp. 4390--4405, 2022.

\bibitem{islam2025}
S.~A.~U. Islam and D.~S. Bernstein, ``Output-feedback model predictive control
  of nonlinear systems with piecewise-linear input-output dynamics,'' in
  \emph{Proc. Amer. Contr. Conf.}\hskip 1em plus 0.5em minus 0.4em\relax IEEE,
  2025, pp. 1876--1881.

\end{thebibliography}

\end{document}